# Why are so many primitive stars observed in the Galaxy halo?

Carl H. Gibson[a1], Theo M. Nieuwenhuizen[b] and Rudolph E. Schild[c]

[a] University of California at San Diego, La Jolla, CA, 92093-0411, USA
[b] Institute for Theoretical Physics, Valckenierstraat 65, 1018 XE Amsterdam, The Netherlands
[c] Harvard-Smithsonian Center for Astrophysics, 60 Garden Street, Cambridge, MA 02138, USA

**ABSTRACT**

Small values of lithium observed in a small, primitive, Galaxy-Halo star SDSS J102915 + 172927 cannot be explained using the standard cold dark matter CDM theory of star formation, but are easily understood using the Gibson/Schild 1996 hydrogravitational-dynamics (HGD) theory. From HGD, primordial H-$^4$He gas fragments into Earth-mass planets in trillion-planet proto-globular-star-cluster (PGC) clumps at the 300 Kyr time of transition from the plasma epoch, soon after the big bang. The first HGD stars formed from pristine, frictionally-merging, gas-planets within the gently stressed clumps of the early universe, burning most available lithium in brown-dwarfs and hot-stars before creating metals that permit cooler burning. The Caffau halo star is a present day example. CDM first stars (Population III) were massive and promptly exploded, re-ionizing the gas of the universe and seeding it with metals, thus making the observed star unexplainable. From HGD, CDM and its massive first stars, and re-ionization by Pop III supernovae, never happened. All stars are formed from planets in primordial clumps. HGD first stars (Pop III) were small and long-lived, and the largest ones were hot. We suggest such small HGD (Pop III) stars still form in the gently stressed Galaxy halo.

**Keywords:** Cosmology, star formation, planet formation, astrobiology.

## 1. INTRODUCTION

The standard model of cosmology ΛCDMHC is in trouble on all sides. It fails to permit life to form by natural causes (Gibson, Wickramasinghe, Schild 2011; Gibson, Schild, Wickramasinghe 2011). It fails to include basic fluid mechanical processes crucial to gravitational structure formation of the various astrophysical fluids; that is, the kinematic viscosity, turbulence and stratified turbulent mixing. It requires non-physical entities such as persistent anti-gravity dark energy Λ and weakly interactive cold-dark-matter CDM particles that clump rather than diffuse in ways that are repudiated by astronomical measurements (Kroupa et al. 2011) and fluid mechanical theory (TMN declines to criticize Λ). CDM clumps have never been convincingly observed, and neither are the numerous small galaxies that are expected within CDM-clump gravitational potential wells as precursors to normal galaxies by hierarchical clustering (HC). ΛCDMHC fails most miserably in its predictions about star and planet formation. As predicted by hydro-gravitational-dynamics HGD cosmology (Gibson 1996), and as observed by quasar microlensing (Schild 1996), for every star in a galaxy there should be 30 million planets, not 8-10 as expected from ΛCDMHC. In particular, the primitive Galaxy-Halo stars and their Lithium abundances observed by Caffau and her colleagues (Caffau et al. 2011) are easily explained and expected from HGD cosmology, but are quite impossible to explain from ΛCDMHC cosmology and its CDM-halo models of star and galaxy formation, which are accepted and taken to be standard in astronomy. The "impossible" Caffau star, as it has come to be known, is shown in Figure 1.

---
[1] Corresponding author: Depts. of MAE and SIO, CASS, *cgibson@ucsd.edu*, **http://sdcc3.ucsd.edu/~ir118**

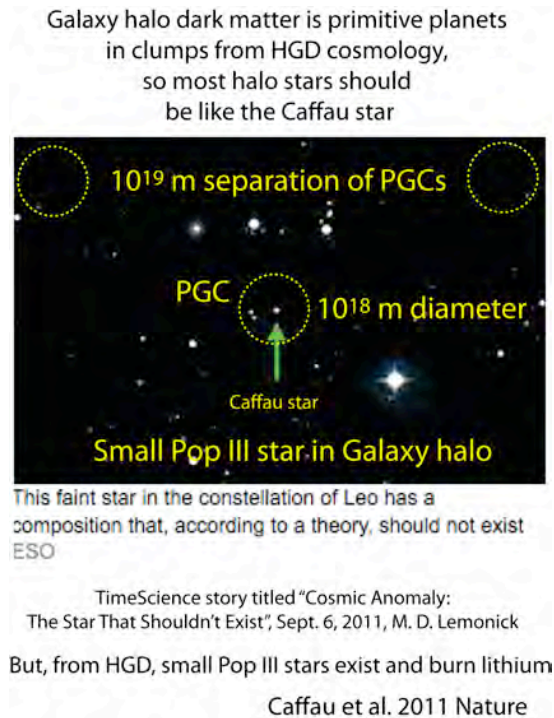

Figure 1. Spectroscopic studies using the Very Large Telescope (VLA) reveal distant Galaxy-Halo stars that cannot be explained using the standard model of cosmology and the standard model of star formation (Caffau et al. 2011). From HGD cosmology, all stars form in PGC clumps of gas planets, the dark matter of galaxies, and burn lithium and tritium while the Pop III stars are small. Small Pop III stars are impossible according to ΛCDMHC cosmology (Bromm and Loeb 2003).

As shown in Fig. 1, the Caffau star appears very ordinary until examined by the large VLT telescope in Chile for its chemical composition using standard (ΛCDMHC) models and the sensitive, precise, and extensive spectroscopic evidence the VLT produces. Using the data and the standard ΛCDMHC model, the Caffau star is impossible.

The difficulty is with the lithium abundance, which is far below levels observed in the Galaxy disk and the expected primordial levels, as shown in Figure 2 (adapted from Fig. 2a of Caffau et al. 2011). The small levels of lithium in halo stars, compared to disk stars, is illustrated using the Spite plateau (Spite & Spite 1982), shown by the light dashed line, which is less than primordial levels (dark dashed line) by a factor of 2-3. Such large differences can no longer be attributed to errors in the primordial lithium abundance. Experts in the field admit that the problem worsens (Cyburt et al. 2008).

A more likely explanation is that Galaxy disk stars have been seeded with metals produced by supernovae that are more frequent for tidally-agitated disk-PGCs compared to pristine gentle-halo-PGCs. This makes a difference from HGD cosmology, where stars are formed from planets in clumps that isolate their supernova chemicals, but does not for ΛCDMHC cosmology, where stars and galaxies condense from gas and dust in >$10^{21}$ m CDM halos that incubate and contaminate internal stars and galaxies, diameters much larger than <$10^{18}$ m PGCs. Galaxy-halo planet-clumps (PGCs) can remain relatively uncontaminated by metals until their first supernova. All their stars burn lithium, starting with brown dwarfs. From HGD, star brightness increases with tidal agitation of the PGC, which increases the planet accretion rate. The more pristine the planets the hotter the largest stars that they create (~0.8 solar in globular star clusters).

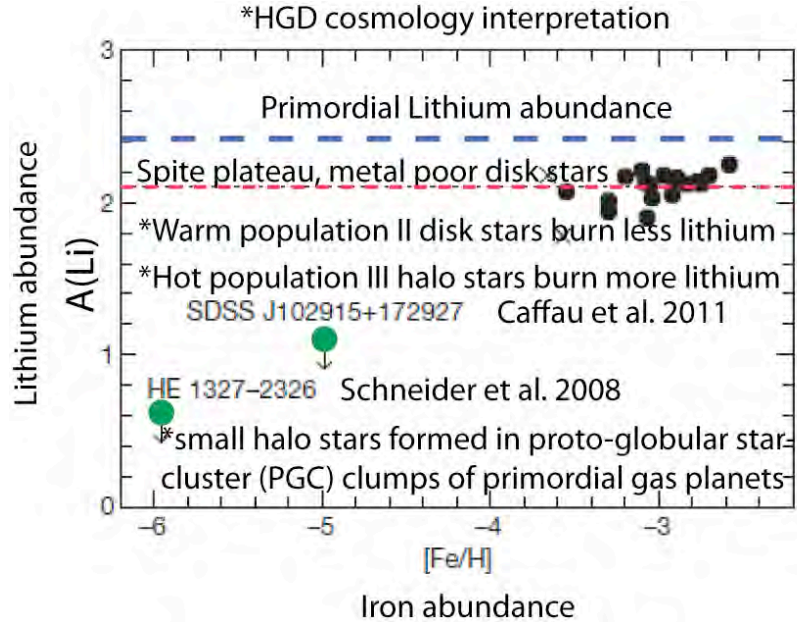

Figure 2. Spectroscopic studies using the Very Large Telescope VLA reveal distant Galaxy-Halo stars that cannot be explained using the standard model of cosmology and the standard model of star formation (Fig. 2a Caffau et al. 2011). The heavy dashed blue line is the primordial gas lithium abundance. The light red dashed Spite plateau (Spite & Spite 1982) is lower by a factor of two, but the Caffau and Schneider halo stars (green) are lower by orders of magnitude. The HGD interpretation (asterisks) is that halo stars in their PGC dark matter clumps are less likely to be seeded by supernovae metals than disk stars in their more tidally agitated, and therefore explosive, PGCs.

As shown in Fig. 2, HGD cosmology gives a straightforward explanation for the small lithium abundances observed (Caffau et al. 2011, Schneider et al. 2008) in small halo stars. Because all stars are formed by planet mergers within PGC clumps, and because PGC clumps in the halo of galaxies are more likely to consist of pristine primordial gas planets that those in the disk, the stars formed are likely to burn all of their lithium during formation rather than the cooler population II stars of the disk that have been seeded with supernovae metals. Because the first ΛCDMHC stars are so large and their supernovae so violent, stars from uncontaminated primordial gas become rare or nonexistent. The abundance of metal A compared to B is found from the expression [A/B] = log($N_A/N_B$) - log($N_A/N_B$)$_{solar}$.

Nieuwenhuizen (2011a,b) examines the effects of $^3$He concentrations on star formation and death, as well as the formation of early galactic central black holes and bulges from Jeans clusters (PGCs) of µBDs (micro-brown-dwarfs). In the following we will discuss the star formation theories of ΛCDMHC versus HGD cosmologies, and make comparison to observations. Conclusions are presented.

## 2. THEORY

Gravitational structure formation in hydro-gravitational-dynamics HGD cosmology depends on kinematic viscosity, turbulence, and molecular diffusivity, quantities that are neglected entirely by ΛCDMHC cosmology. The two cosmologies give very different predictions with respect to the formation of stars and the formation of planets. How do these predictions affect the interpretation of the Caffau star? What is the evidence supporting the two cosmologies?

Cold Dark Matter is in trouble for several reasons. The only known non-baryonic dark matter material is neutrinos, and the most popular alternative candidate, the neutralino, is failing all tests using the Large Hadron Collider, as shown in Figure 3.

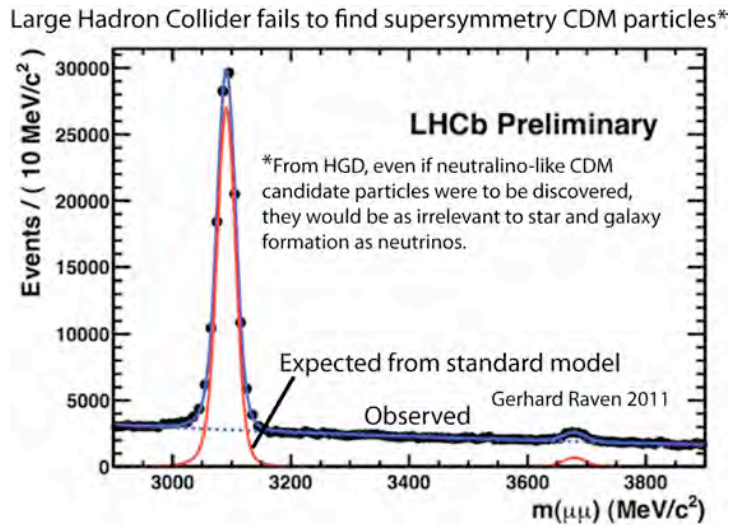

Figure 3. Preliminary results from the Large Hadron Collider show departures of the Observed radiation measured by the LCHb detector (blue) compared to that expected using the standard model of particle physics (red). The Figure is taken from a slide presented (Raven 2011) in August at the Lepton-Photon particle physics meeting in Mombai, India.

Neutralinos are predicted by supersymmetry as cold dark matter particles because they are massive, about a hundred times a proton mass, and weakly collisional, with expected collision cross sections as small as $10^{-49}$ m$^2$. The LCHb strategy is to look for collisions of bottom quark related particles with massive supersymmetry CDM candidates. The preliminary results reported by Gerhard Raven (LP meeting Mombai 2011) so far show no evidence that such particles exist.

Star formation by the standard ΛCDMHC cosmology (Bromm and Loeb 2003) is very different than HGD star formation. CDM seeds that form by the Jeans 1902 instability theory somehow hierarchically cluster (HC) to form CDM halos of larger mass. Stars form as the primordial gas falls into the resulting potential wells. A period of 400 Myr or more called the "dark ages" is required before any stars appear. Gravity pulls all the gas to the center of the CDM halo, so the first stars were enormous, metal free, Population III stars that rapidly exploded to form metals. The supernova were so powerful they completely re-ionized and contaminated all the gas of the universe. Attempts to detect this "first starlight" have failed, probably because it never happened (Madau 2006). ΛCDMHC assumes that a threshold mass fraction of metals $Z > 10^{-7}$ is required to form stars smaller than 0.8 solar mass (Schneider et al. 2003). HGD claims star brightness shows the rate of planet accretion, not the star mass, and that Z is irrelevant to star brightness. The first stars were small Population III stars of the old globular star clusters (OGCs).

All ΛCDMHC stars after re-ionization should be contaminated by metals from the supernovae. They should burn cooler than the small Population III stars that are expected in the Galaxy halo from HGD (Fig. 1). It is a mystery to ΛCDMHC how any small Population III stars could be formed (Bromm & Loeb 2003), let alone the very large numbers of small Population III halo stars observed, such as the Caffau star of Fig. 1 and Fig. 2 (Frebel et al. 2008).

The mystery is easily solved by HGD cosmology (Gibson 1996, Schild 1996). Rather than cold dark matter condensation during the plasma epoch, structure formation begins when the viscous forces match the gravitational forces at the Schwarz viscous scale $L_{SV} = (\gamma \nu / \rho G)^{1/2}$, where $\gamma$ is the rate-of-strain, $\nu$ is the kinematic viscosity, $\rho$ is the density and G is Newton's gravitational constant. Structure formation is prevented by the photon viscosity of the plasma until time $10^{12}$ seconds, when $L_{SV} > L_H$ first matches the increasing scale of causal connection $L_H = ct$, where c is the speed of light and t is the time. The computed mass of the first structures found in this way closely matches the observed mass of superclusters (Gibson 2000), about $10^{46}$ kg.

The last structures formed in the plasma are protogalaxies, with mass about $10^{43}$ kg and linear morphology caused by fragmentation along vortex lines of weak turbulence produced by expanding protosuperclustervoids (Gibson, Schild and Wickramasinghe 2011). Rather than condensing into CDM halos, the non-baryonic dark matter is super-diffusive. It forms the halos of clusters and superclusters and a negligible part of galaxy mass. A substantial portion of the non-baryonic dark matter appears to be neutrinos (Nieuwenhuizen & Morandi 2011) from gravitational lensing by a galaxy cluster. A permanent form of anti-gravity (dark energy) is not needed by HGD cosmology (TMN reserves judgement), since big bang turbulence supplies the large anti-gravity negative stresses ($10^{113}$ Pa) required for mass-energy extraction at Planck scales, by vortex stretching (Gibson 2010).

The phase transition from plasma to gas occurs at $t = 10^{13}$ seconds. Because heat is transferred at the speed of light and pressure at the speed of sound, the protogalaxies fragment at two length scales: the Jeans scale $L_J = V_S \tau_g$ and $L_{SV}$, where $V_S$ is the speed of sound and $\tau_g = (\rho G)^{-1/2}$ is the gravitational free fall time. Each protogalaxy fragments into $10^{18}$ m Jeans mass clumps (PGCs) of primordial gas planets (now frozen-solid-hydrogen $10^7$ m µBD microBrownDwarfs) that have persisted as the dark matter (Nieuwenhuizen, Schild and Gibson 2011). As the universe cooled, more and more planets froze and their PGCs diffused from the central core to form the galaxy halo and galaxy accretion disk. Most of the PGCs of the halo remain as clumps of a trillion, pristine, primordial-gas-planets in metastable equilibrium, with no stars whatsoever. Whatever stars form are most likely to be similar to the Caffau star of Fig. 1 and Fig. 2, with the small Pop III stars of old globular clusters. There should be many such stars, with very low lithium abundance, as observed (Sbordone et al. 2010).

In a future work we will show how the long-standing ΛCDMHC mystery of $^7$Li abundance (Cyburt et al. 2008) can be addressed using HGD cosmology.

### 3. CONCLUSIONS

Large numbers of primitive halo stars revealed by the VLT (Very Large Telescope) and its very sensitive spectrographs are easy to understand using HGD cosmology, but are impossible to explain using standard ΛCDMHC cosmology. The reason is that the standard cosmology is wrong about how stars are formed. Stars are not formed from gas that falls into CDM gravitational potential wells, they are formed within proto-globular-star-cluster (PGC) clumps of primordial gas planets, by mergers of the planets to form larger planets and finally stars. PGCs that freeze and diffuse into the galaxy halo are generally pristine, with no stars at all and no metals besides traces of lithium. Their planets merge to form larger planets, brown dwarfs and hot small stars that burn all traces of primordial lithium. Lithium abundances in Population II disk stars are only a factor of three or less smaller than primordial gas values, compared to more than a factor of ten for many halo stars observed.

No observations support the existence of CDM halos. The LHC experiments of Fig. 3 show no evidence that the leading CDM particle candidates exist. Even if they did, they would be irrelevant because they are weakly collisional and would simply diffuse away from the baryons, just as neutrinos do. A host of new observations, including those of the Caffau star, suggest it is time to abandon ΛCDMHC cosmology in favor of HGD.